\documentclass[amsmath,twocolumn,superscriptaddress]{revtex4-1}
\usepackage{epsfig}
\usepackage{amssymb}
\usepackage{amsmath}
\usepackage{amsfonts}
\usepackage{braket}
\usepackage[T1]{fontenc}
\usepackage{bm}
\usepackage{graphicx}
\usepackage{xcolor}
\usepackage{verbatim}
\usepackage{soul}
\usepackage{tikz}
\usetikzlibrary{matrix,shapes,arrows,positioning,chains}
\usetikzlibrary{shapes.geometric}
\usetikzlibrary{shapes.arrows}
\usepackage{array}
\usepackage{enumitem}
\usepackage{color}
\usepackage{physics}
\usepackage[export]{adjustbox}
\usepackage[percent]{overpic}
\usepackage{filecontents}

\begin{document}
\title{Phase Transitions in the Classical Simulability of Open Quantum Systems}

\author{F. Azad}
\affiliation{London Centre for Nanotechnology, University College London, Gordon St., London, WC1H 0AH, United Kingdom}

\author{A. Hallam}
\affiliation{School of Physics and Astronomy, University of Leeds, Leeds LS2 9JT United Kingdom}

\author{J. Morley}
\affiliation{London Centre for Nanotechnology, University College London, Gordon St., London, WC1H 0AH, United Kingdom}

\author{A.~G. Green}
\affiliation{London Centre for Nanotechnology, University College London, Gordon St., London, WC1H 0AH, United Kingdom}

\date{\today}
\begin{abstract}
We study the evolution of an open quantum system using a Langevin unravelling of the density matrix evolution over matrix product states. As the strength of coupling to and temperature of the environment is increased, we find a transition where the entanglement of the individual trajectories saturates, permitting a classical simulation of the system for all times. This is the Hamiltonian open system counterpart of the saturation in entanglement found in random circuits with projective or weak measurements. If a system is open, there is a limit to the advantage in simulating its behaviour on a quantum computer, even when that evolution harbours important quantum effects. 
\end{abstract}
\maketitle
%\tableofcontents
 
The term classical is applied to quantum systems in at least two different ways. On the one hand, if a closed quantum system is in a weakly-entangled state, it may be considered classical as long as the entanglement remains low. In this limit, the equations of motion of the system are termed semi-classical. On the other hand, an open system behaves classically once the coupling to the environment has caused dephasing of the off-diagonal elements of the density matrix. These two limits occur on very different timescales --- the semi-classical limit at early times and the dephasing limit at late times. 

Can these views be reconciled and a classical description developed that works from the earliest to the latest times? Recent insights have made steps towards such an understanding for open systems. The transition in entanglement growth in random circuits as a function of the rate of projective or weak measurement allows a classical, weakly-entangled description of the system for all times\cite{nahum2017quantum,li2018quantum,nahum2018operator,von2018operator,chan2019unitary,li2019measurement,szyniszewski2019entanglement}. 

Both are cousins to the quantum Zeno effect, by which frequent measurement in a channel impedes transitions in that channel\cite{misra1977zeno}. The nature of the many-body transition has been studied extensively\cite{szyniszewski2020universality,li2020conformal,chen2020emergent,tang2020measurement,zhang2020nonuniversal,zabalo2020critical,gullans2020scalable,gullans2020dynamical,jian2020measurement,choi2020quantum,bao2020theory} and similar analyses extended to measurement-induced transitions in open Hamiltonian systems\cite{goto2020measurement,alba2021spreading,fuji2020measurement,PhysRevLett.126.170602}.  These later cases are closely related to a transition in classical describability as a function of coupling to the environment, which we consider.

Our approach is to consider an unravelling of the density matrix evolution for an open system over trajectories described using matrix product states. The equations of motion of each trajectory can be considered a Langevin extension of the time-dependent variational principle (TDVP) for matrix product states. 
For closed systems, the TDVP equations constitute a semi-classical limit; they correspond to classical Hamilton equations of motion on the variational manifold\cite{HaegemanTDVP,hallam2019lyapunov}. As the entanglement grows during the Hamiltonian evolution, the TDVP equations break down as a larger and larger variational manifold (measured by an exponentially growing bond order) is required in order to capture the state and its dynamics. In this sense, the semi-classical description is confined to early times. In our stochastic TDVP Langevin equation, we find thresholds in the dynamics of individual trajectories as a function of coupling to and temperature of the environment, whereby the entanglement saturates at a low value, and a low-bond order description gives high fidelity for all time. 
This {\it quantum Zeno phase}\cite{li2018quantum} constitutes a transition in the classical describability of the open quantum system: the low bond order TDVP Langevin equation is an effective semi-classical description that works for all time.

\section{A TDVP Langevin Equation} 
\label{Sec:MPS Langevin Equation}
%%%%%%%%%%%%%%%%%%
%
\begin{figure}[!t]
%b) \hspace{2in}\; \\
\includegraphics[width=0.5\textwidth]{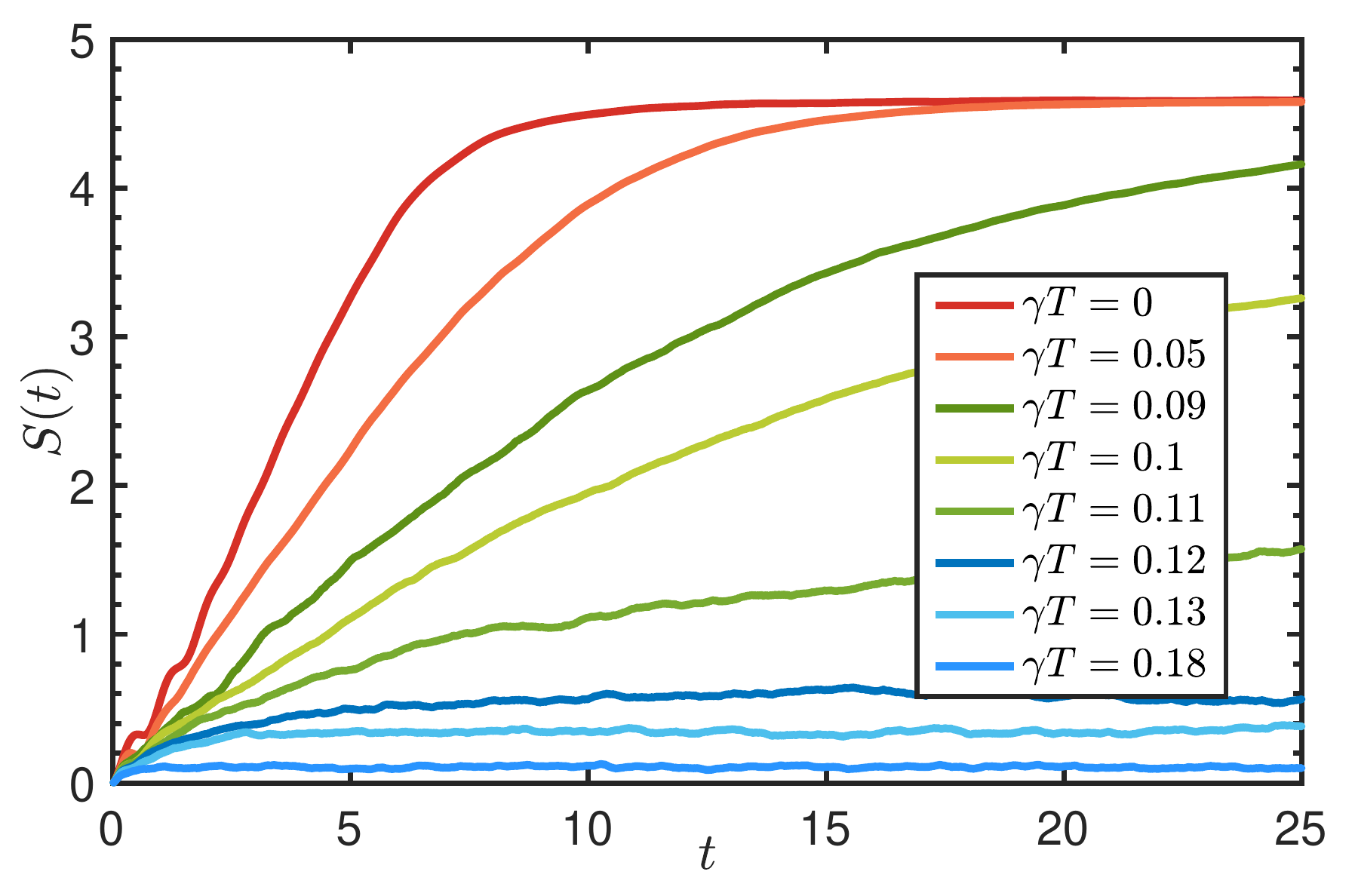}
\caption{
{\it Basic Properties of MPS Langevin Evolution:} The TDVP Langevin equation, Eq.(\ref{eq:TDVPLangevin}) describes the evolution of the density matrix through an ensemble of stochastic pure-state trajectories.  
%a) At finite temperature with weak friction, the late time evolution is towards a thermal ensemble on the variational manifold. The figure shows the energy {\it versus} temperature for a fixed, small value of $\gamma=0.05$  under Eq.(\ref{eq:TDVPLangevin}) with Hamiltonian Eq.(\ref{eq:Hamiltonian}) and $J=1$, $g=-1.05$ and $h=0$, compared to a direct thermal average computed over the variational manifold [See Appendix \ref{app:HaarAverage}]. The thick lines correspond to a fitting to the thermal average, while the crosses in black are corresponding energy values for the Langevin case. 
Here, the von Neumann entanglement as a function of time for a typical trajectory is shown calculated is shown calculated for different coupling strengths. Here the bond dimension is $D=128$ and we have kept temperature fixed, $T=0.2$, while we vary the noise, $\gamma T$, i.e. coupling $\gamma$. Such plots show three different regimes of behaviour. An initial transient, followed by an approximately linear in $t$ or logarithmic in $t$ growth and finally a saturation at the longest times. This saturation can either be determined by the variational approximation or intrinsically by the interplay of the Hamiltonian and the dissipative bath. It signals whether a lower bond-dimension (hence less computationally intensive) simulation suffices in underpinning the entropy dynamics.}
\label{fig:BasicProperties}
\end{figure}

Langevin equations describe the motion of a system coupled to an environment (or alternatively the motion of slow collective degrees of freedom in an effective bath described by the faster degrees of freedom\cite{zwanzig1960ensemble}) by adding noise and friction terms to the basic equations of motion of the system. If the environmental degrees of freedom are in thermal equilibrium, the friction and noise satisfy a fluctuation-dissipation relation. Applied to quantum systems, the Schr\"odinger equation provides the basic equations of motion. The ensemble of the resulting stochastic Schr\"odinger trajectories recovers the density matrix evolution and is said to be an unravelling of it. 

{\it The Langevin equation} over matrix product states studied here, can be written in its Markovian limit as:
\begin{eqnarray}
\langle \partial_i \psi | \partial_j \psi \rangle \dot X_j
&=&
-i \langle \partial_i \psi | \hat H | \psi \rangle
-i \sum_n \langle \partial_i \psi | \hat F_n | \psi \rangle \eta(t)
\nonumber\\
& &
-i \sum_n \gamma \frac{\langle \psi | \hat F_n | \psi \rangle }{dt}  \langle \partial_i \psi | \hat F_n  | \psi \rangle .
\label{eq:TDVPLangevin}
\end{eqnarray}
The terms on the left-hand side and the first term on the right constitute the conventional time-dependent variational principle (TDVP) equations\cite{HaegemanTDVP}. The second and third terms on the right are, respectively, the noise and friction due to coupling to the environment. $\hat F_n$ are the operators by which the system is coupled to the bath displacement operators. We generally assume these to be spatially local. For spin-half chains they are given by the $x,y $ and $z-$components of the spin operators on each site, each of which couples to a separate bath. The bath is described as a collection of harmonic oscillators and the noise-correlator is determined by the spectrum of oscillators and the temperature of the bath; $\langle \langle \eta(t) \eta(t') \rangle \rangle=2 \gamma T \delta(t-t')$ in the combined classical and Markovian limits. 

{\it The TDVP Langevin Equation} is a new approach to unravelling the density matrix evolution of an open system. It combines the study of quantum many-body dynamics on a variational manifold --- here we use matrix product states (MPS)\cite{Orus:2014zl,schollwock2011density,perez2006matrix} --- with the Langevin limit of Keldysh field theories\cite{kamenev2011field,sieberer2016keldysh}. 
We combine these approaches by constructing a Keldysh path integral over MPS states\cite{green2016feynman}. To date, MPS techniques have been employed in the study of open systems largely in three ways: by starting with the Lindblad master equation and either describing the density matrix directly as a matrix product operator\cite{verstraete2004matrix,cui2015variational,weimer2019simulation} or else unravelling its evolution over MPS representations of quantum trajectories\cite{daley2009atomic,daley2014quantum,bonnes2014superoperators}; alternatively, tensor network techniques can be employed to describe the bath directly within the state ansatz\cite{strathearn2018efficient}. Of these, the latter is numerically costly and the Lindblad-based approaches formally apply in a different limit\cite{FergusThesis}. 
Questions of applicability aside, there are some features of our approach that make it particularly attractive.
It naturally leads to thermal equilibrium, and it can treat the non-Markovian limit. There also exists a natural hydrodynamic limit and Fokker-Plank description. The scaling of the algorithm for each trajectory is the same as the usual TDVP for matrix product states -- moreover, there appears to be a degree of self averaging so that certain quantities are well-approximated by relatively fewer trajectories. 

{\it A derivation} of Eq.(\ref{eq:TDVPLangevin}) is given in Appendix A. We develop the Langevin equation from the Keldysh path integral for the time-evolution of the density matrix. The method follows that of Ref.\cite{kamenev2011field} with the modification that the Keldysh path integral is constructed over matrix product states\cite{green2016feynman}. The result adds noise and friction to the time-dependent variational principle constructed over matrix product states\cite{HaegemanTDVP}. A similar construction for any variational class would lead to a similar stochastic equation of motion, which we dub the TDVP-Langevin equation. An alternative, heuristic derivation involves solving the Schr{\"o}dinger equation for the system and bath and substituting the solution of the latter into the equation of motion of the former. Alternative unravellings of the Lindblad equation for the density matrix evolution over MPS\cite{daley2009atomic} apply in different circumstances of relative time and energy scales of the bath and system. 

We construct Eq.(\ref{eq:TDVPLangevin}) over matrix product states using conventional methods\cite{HaegemanTDVP}. Integration of this equation is complicated by the friction term. Naively, this requires inversion of a matrix that is proportional to the system size and dimension of the variational manifold. However, recognising that it consists of an outer product of vectors allows an efficient inversion and integration of the equations of motion. Details are given in Appendix B and our code is available at https://github.com/AndrewHallam/Langevin.

{\it Basic properties} of the TDVP Langevin equation are summarised in Fig.\ref{fig:BasicProperties}.  A
 thermal distribution over the variational manifold is given by a Boltzmann-weighted Haar average over the variational manifold. In the case the of MPS of bond dimension $D$, this average can be performed as a Haar integral over the group $SU(dD)$ (with $d$ the local Hilbert space dimension). Such thermal distributions are fixed points of the Langevin evolution (see Appendix \ref{app:HaarAverage}). 
Fig.~\ref{fig:BasicProperties} captures the dynamics of the von Neumann entanglement entropy typical for trajectories over the parameters we consider. In the case shown, temperature $T=0.2$ is kept fixed and friction is increased. Beyond a critical value, $\gamma T\approx 0.1$, entanglement growth is suppressed. The entanglement undergoes a transition from being determined by the variational approximation to becoming intrinsic to the interplay between the Hamiltonian and the dissipative bath. Similar transitions occur when other parameters are kept constant (friction  $\gamma$, or noise  $\gamma T$). We detail these results in the following section.

\section{Open Evolution of a Rapidly Entangling System}
\label{Sec:Results}
%%%%%%%%%%%%%%%%%%%%%%%%%%%%
In the absence of coupling to a bath, TDVP equations eventually fail as the entanglement grows beyond that which can be represented on the variational manifold\footnote{TDVP equations for the thermofield purification of the density matrix may escape this fate\cite{hallam2019lyapunov} at least as far as local observations are concerned}. However, just as observed in projective measurements of random circuits, the effects of the environment may restrict the growth of entanglement. {\it In extremis} this might limit entanglement of individual trajectories so that they can be represented on low dimensional variational manifolds. The TDVP Langevin equation will then give a good account of the dynamics at all times, signifying a transition in its classical representability. This is our interpretation of the sequence of results presented in this section.

{\it The Hamiltonian} that we consider is the tilted field Ising model 
\begin{equation}
\hat H =
-\sum_i \left[ J \sigma_i^z \sigma_{i+1}^z +h \sigma^z_i+ g \sigma^x_i \right],
\label{eq:Hamiltonian}
\end{equation}
with $J=1$, $g=-1.05$ and $h=0.5$. With these parameters, the Hamiltonian is a far from any integrable point and rapidly thermalising\cite{banuls2011strong,leviatan2017quantum}.

\begin{figure}
\includegraphics[width=0.47\textwidth]{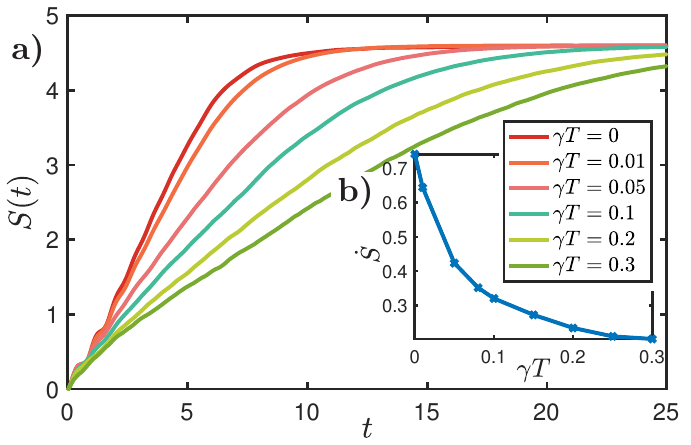}\caption{
{\it Evolution of Entanglement at Infinite Temperature:} Here we consider the evolution under Eq.(\ref{eq:TDVPLangevin})  with Hamiltonian Eq.(\ref{eq:Hamiltonian}) with $J=1$, $g=-1.05$ and $h=0.5$. The friction coefficient $\gamma=0$ with noise $\gamma T$ finite and initial state $Z$.
 a) von Neumann Entropy {\it versus} time at a fixed bond order $D=160$ for different values of $\gamma T$. Without noise, and after an initial transient, the entanglement grows linearly with time before saturating at a constant value less than the maximum determined by the MPS manifold. With non-zero noise, the entanglement growth is linear in time with a growth rate that reduces with increasing $\gamma T$ due to the dephasing effects of the bath. Ultimately, all curves saturate (not shown) to the same value characteristic of an infinite temperature state on the variational manifold. 
b) The growth rate of entanglement $\dot{S}$, extracted beyond the initial transient ($t\approx4$), {\it versus} noise.} 
\label{fig:InfiniteT}
\end{figure}

{\it Infinite temperature and vanishing friction}: Fig.\ref{fig:InfiniteT} shows the variation in von Neumann entanglement across the central bond as a function of time for simulations with a range of bond orders and noise strengths. The broad result of these simulations is that the entanglement saturates at long times at a value determined by the bond order of the simulation. This is consistent with an infinite temperature final state with the maximum entanglement supported by the variational manifold. The most interesting aspect of these results is the decreasing rate of early-time entanglement growth with increasing noise strength. Crucially we do not find evidence of an intrinsic saturation of entanglement -- only that dictated by the limitations of the variational approximation.

\begin{figure*}
%\centering
\includegraphics[trim={6 0 0 0},clip, width=0.32\textwidth]{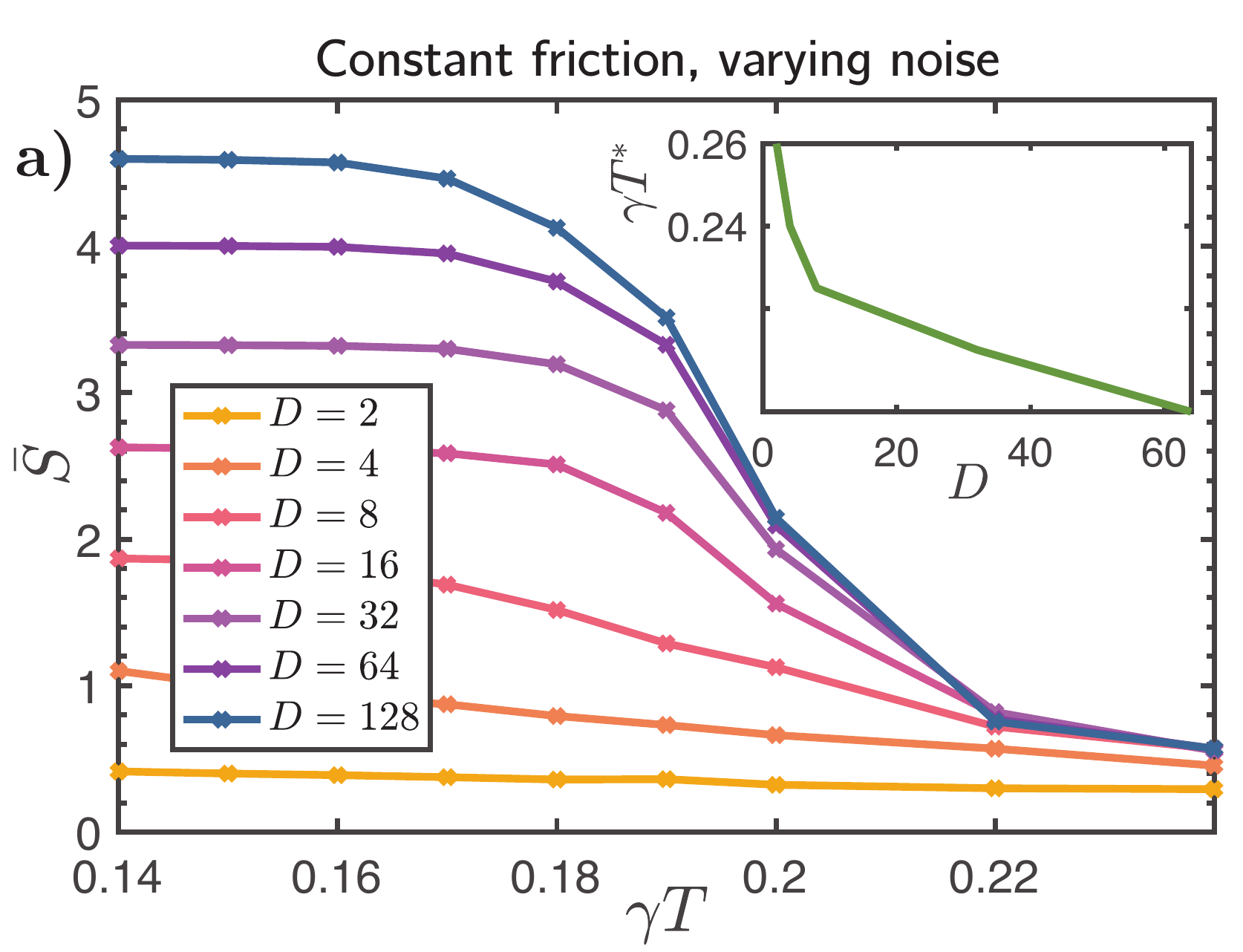}
\includegraphics[trim={5 4 5 0},clip, width=0.32\textwidth]{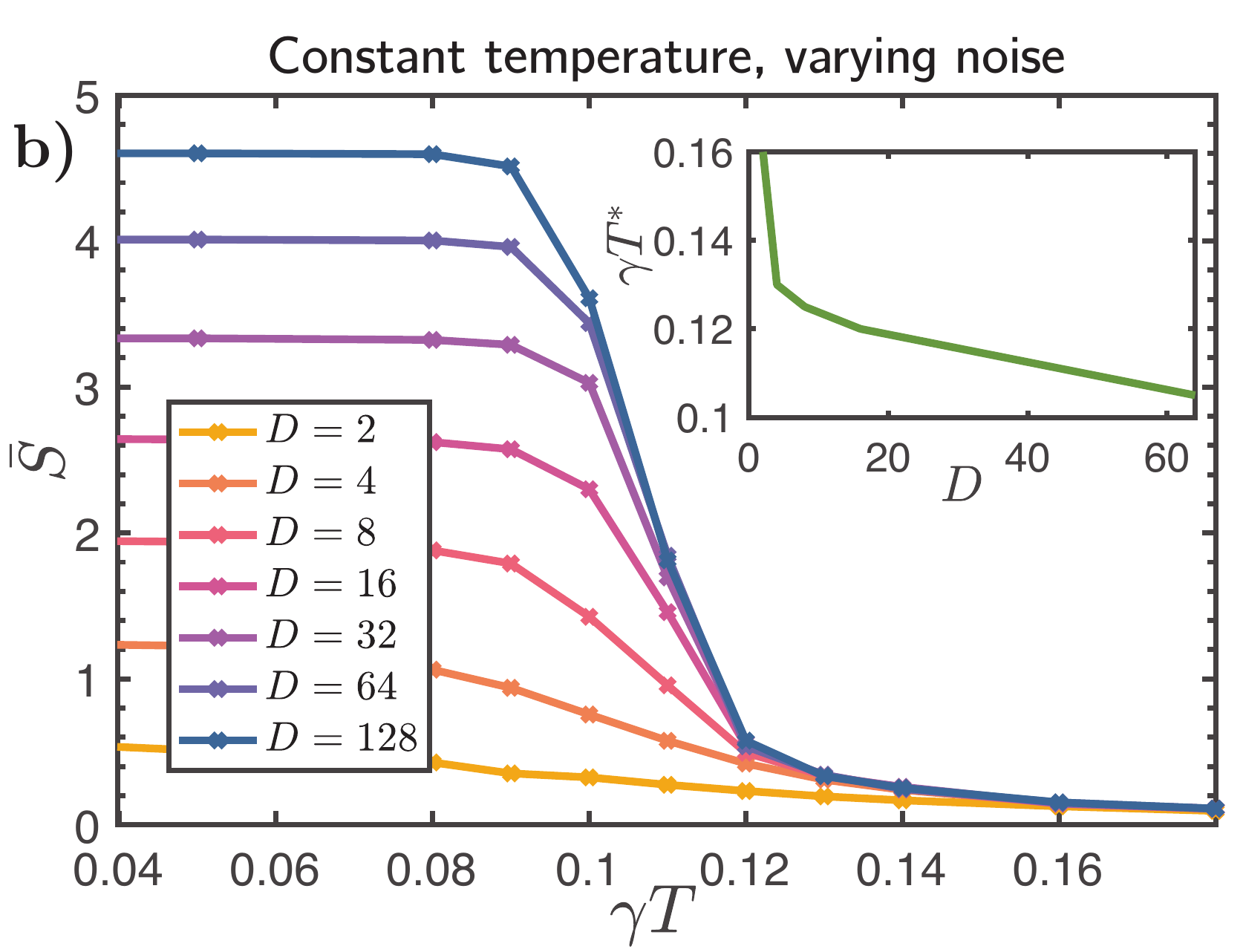}
\includegraphics[trim={0 0 6 0},clip, width=0.32\textwidth]{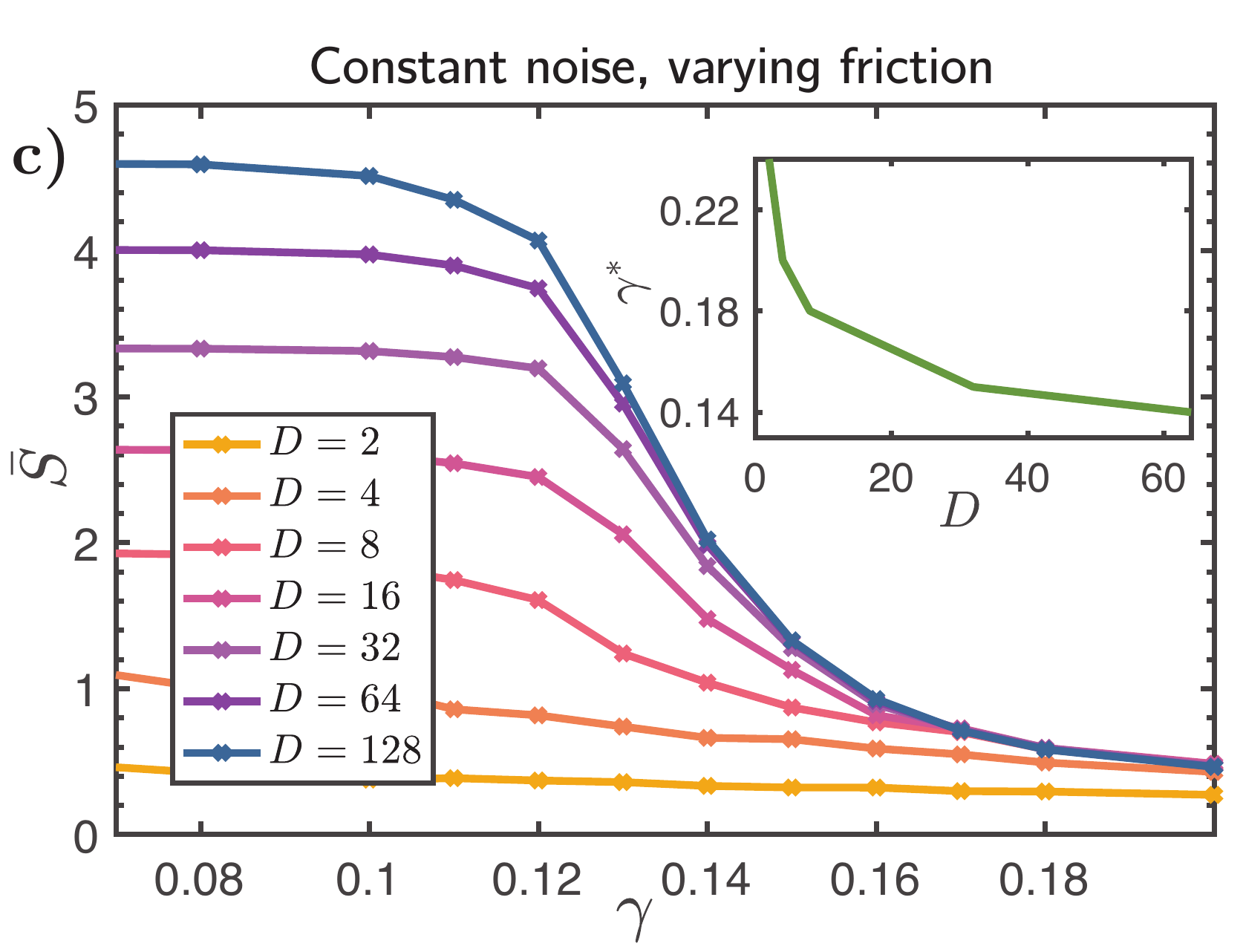}
\caption{
{\it Evolution of Saturation Entanglement at Finite Temperature:} Here we consider the evolution under Eq.(\ref{eq:TDVPLangevin}) with Hamiltonian Eq.(\ref{eq:Hamiltonian}) where $J=1$, $g=-1.05$ and $h=0.5$ and finite $\gamma$ and $T$. In the main figures we show the dependence of the von Neumann entropy as a function of noise and friction: a) {\it versus} $\gamma T$ at fixed $\gamma$, b) {\it versus} $\gamma T$ at fixed $T$, c) {\it versus} $\gamma$ at fixed $\gamma T$. 
In each case, at low values of noise and friction, the saturation entanglement $\bar{S}$, is determined by the choice of variational parametrization through the bond order. As the noise and friction are increased, there is a cross-over where the saturation entanglement decreases from this value. Each bond order captures the saturation entanglement for a sufficiently large noise and friction. This is indicated when the entanglement begins to follow the entanglement given by the highest bond order simulation. Where the saturation entanglement of the individual trajectories (distinguished by bond order) dramatically fall to a single curve, so that the lower bond dimension curves coincide with higher bond dimensions, illustrates this transition to situations where the saturation entanglement is intrinsically determined. {\it In each corresponding inset figure}, we have extracted critical dissipation strengths where these transitions occur as a function of bond order.
}
\label{fig:SaturatingS}
\end{figure*}

\begin{figure*}
\includegraphics[trim={0 0.2cm 0 0},clip,width=17.5cm]{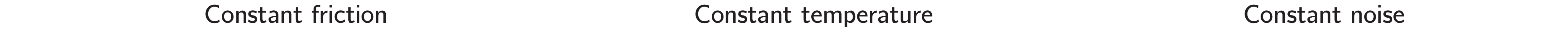}
\includegraphics[trim={0 0 0 0.2cm},clip,width=17.5cm]{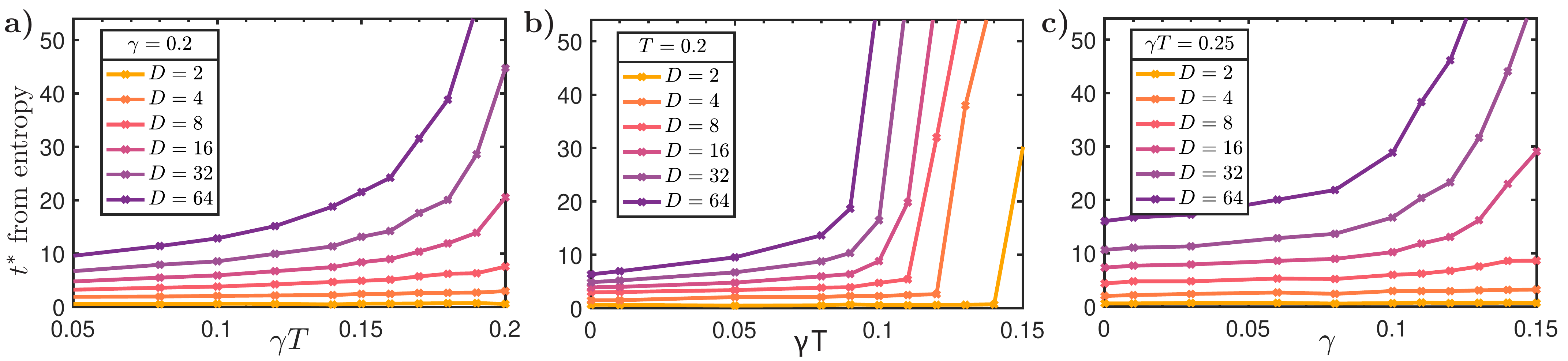}
\newline
\includegraphics[width=17.5cm]{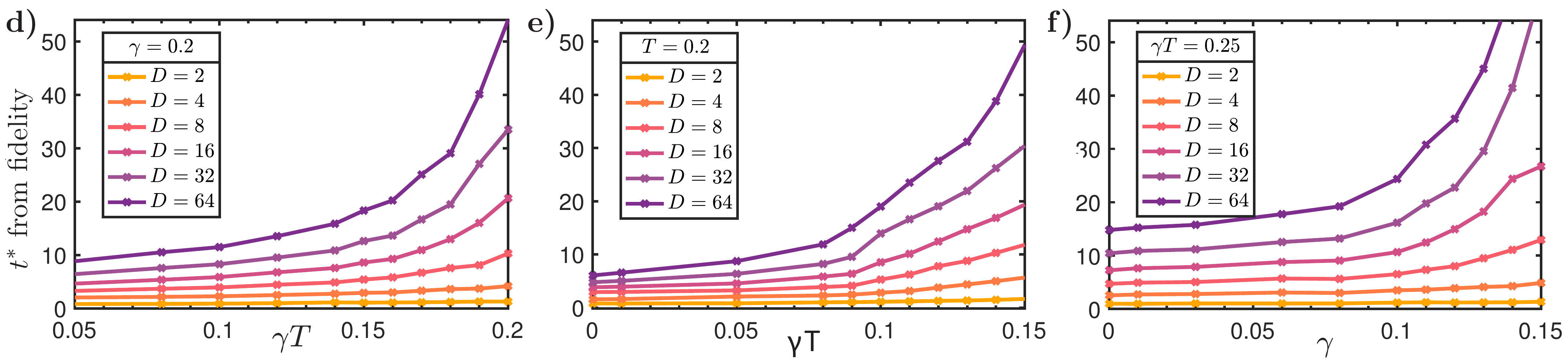}
\newline
\newline
\includegraphics[width=17.5cm]{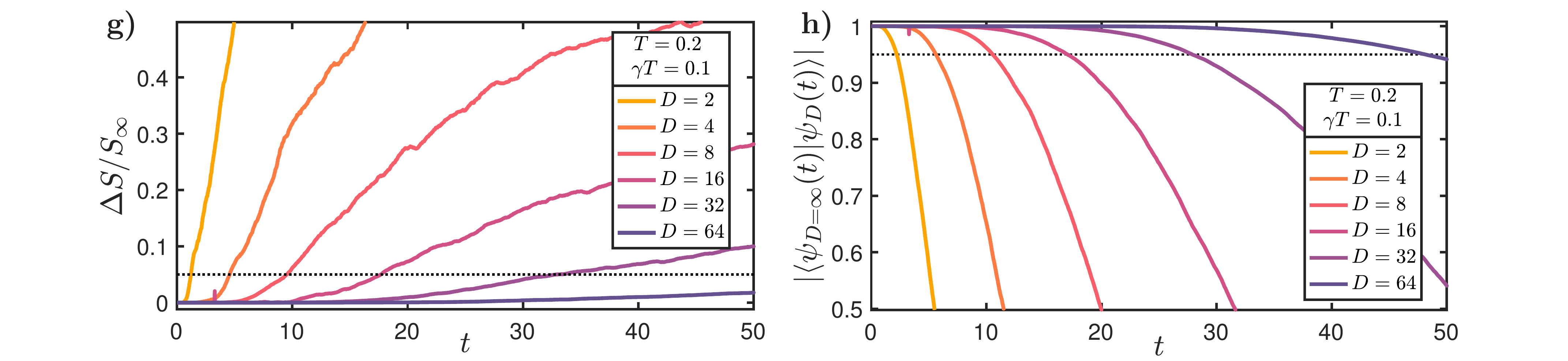}
\caption{
{\it Divergent Classical Simulation Time:} Simulations carried out at a bond order $D$ give a good account of the system evolution up to a time $t^*(D)$. We extract these values versus a reference $D=128$ simulation, which serves as the good account of the system. We do this in two ways, by comparing the difference in von Neumann entanglement entropy between these states and the fidelity with this state. $t^*(D)$ is the time when the simulation with varying bond dimension deviates appreciably from the reference trajectory. The row with panels a), b), c), shows $t^*(D)$ extracted from the entropy, while panels d), e), f) demonstrate this for the fidelity. The fixed variables are split across the columns -- a), d), shows varying $\gamma T$ at fixed $\gamma=0.2$, b), e) $\gamma T$ at fixed $T=0.2$, and c), f) $\gamma$ at fixed $\gamma T=0.25$. Panels g) and h) show typical evolution of entanglement and fidelity with time. This is data from the near-critical point for $T=0.2$, which is for the trajectory with noise $\gamma T = 0.1$. We say that a simulation has failed to provide a good account of the system when the trajectory deviates beyond $\epsilon =0.05$, and the time at which this occurs is $t^*(D)$. In g), this is the point where $\Delta S/S_{D=128} = |S_{D=128}-S_{D}|/S_{D=128} > \epsilon$. Analogously in h), $t^*(D)$ is the time when the fidelity is appreciably different to $1$, i.e. $|\langle\psi_{D=128}(t)|\psi_{D}(t)\rangle|< 1-\epsilon$.
A divergent $t^*(D)$, within either method of extraction, indicates a transition in the classical simulability of the open quantum system.}
\label{fig:DivergentClassicalSimulation}
\end{figure*}

{\it Finite temperature and friction}: 
Including both noise and friction, we do see such an intrinsic saturation. This is demonstrated in two ways; by considering the saturation of entanglement at long times and by a high fidelity between low- and high-bond order simulations at long times. 
\\
\noindent
i. {\it Saturating entanglement}
In order to demonstrate this, we first show in Fig.~\ref{fig:SaturatingS} the long-time average of the von-Neumann entropy. A graph showing the typical time-dependence from which such saturation values are computed is shown in Fig.\ref{fig:BasicProperties}b).  For low noise and friction, the saturation is determined by the limitations of the variational manifold. Panels a), b) and c) show that as a function of $\gamma T$ at fixed $\gamma$, then $T$, and $\gamma T$ at fixed $\gamma$, respectively. A threshold is reached for each bond order where it adequately captures the saturation entanglement, thus indicating a transition to increasingly classically simulatable dynamics. The transition can be seen from the point where the trajectories obtained at different bond orders give the same saturation entanglement. From this we can extract a critical $\gamma$ or $\gamma T$ as a function of bond order that we show in each corresponding inset figure.
\\
\noindent
ii. {\it High fidelity as } $t \rightarrow \infty$: We can identify an analogous transition in the fidelity of each trajectory at different bond orders versus a reference trajectory with bond order $D=128$. In this case, we find that beyond a critical combination of $\gamma$ or $\gamma T$, the fidelity of the state at low bond dimension remains close to $1$ for long times. We expand upon this result in Fig.~\ref{fig:DivergentClassicalSimulation}, where we identify a divergent classical simulation time. We note that the fidelity is more sensitive to the time-step as friction is increased -- an issue typical of numerical integration of systems of stochastic differential equations. This makes accessing the critical point of the transition numerically intensive for the parameters and Hamiltonian we consider. The entropy is less sensitive to this.

\section{Discussion}
%%%%%%%%%%

This work introduces a new method to investigate the dynamics of open many-body quantum systems, the TDVP-Langevin equation. We derive this by considering an appropriate limit of the Keldysh path integral constructed over the MPS manifold. Our investigations reveal a phase transition in the applicability of this approach as a function of coupling to the environment -- when the bath temperature and induced friction are sufficiently high, entanglement growth in individual trajectories is suppressed, and a low bond order description works for all time. This is a transition in the classical simulabiity of the open quantum system. 

We believe that this transition is related to several other transitions in quantum dynamics that have been observed as a function of coupling the the environment or measurement, including the restriction of entanglement growth in random circuits with projective measurement, the quantum Zeno effect (and perhaps the KT transition in the spin-boson model \cite{leggett1987dynamics,Florens2010quantum,barratt2020dissipative}).

The implications of this result may be far-reaching. In the context of using the TDVP-Langevin equation to simulate open quantum systems, an efficient description for long times is possible for systems in the many-body quantum Zeno phase. Indeed, when a target system is in such a phase, there is no (asymptotic) advantage in using a quantum computer to simulate it. Since many chemical reactions of potential interest for quantum computation occur embedded in a dissipative aqueous environment, this is certainly a point worthy of consideration. 

Moreover, viewed from the perspective of a description of the quantum computational device, the transition into the many-body Zeno phase might indicate transitions in the ability to solve quantum problems.  
While thresholds of noise for quantum error correction have been identified in the case of gate-based quantum computation, no such thresholds currently exist for adiabatic computation. It is intriguing to speculate that determining whether a putative adiabatic computational device is in its Zeno phase or not might provide similar bounds on performance\cite{crowley2014quantum,barratt2020dissipative}. 

We envisage a number of ways in which this work might be developed.
Extending the approach to local observables in closed quantum systems presents some exciting possibilities. In this case the bath would refer to other elements of the system itself and its properties self-consistently determined through the evolution\cite{zwanzig1960ensemble,zwanzig1961memory},
Such a description has the promise of connecting early-time semi-classical descriptions to late-time hydrodynamics and thermalisation. Exploring the Fokker-Planck formulation of the TDVP-Langevin equation would bring a complementary perspective to our analysis\cite{berta2017thermal}.

The accurate description of a quantum system from early to late times is generally not possible because of growing entanglement. However, coupling to the environment can limit this growth and render this achievable. This work has coordinated physical insights from several different perspectives to develop such a numerical scheme. We hope both  that the algorithm itself will prove useful and that it will inspire further insights. 

%%%%%%%%%%%%%%%%%%%%%%%%%%%%%%%%%%%%%%%%%%%%%%%
\section{Acknowledgements}
%%%%%%%%%%%%%%
We gratefully acknowledge funding from the EPSRC under grants EP/L015242/, EP/S005021/1 and EP/R020612/1.

%%%%%%%%%%%%%%%%%%%%%%%%%%%%%%%%%%%%%%%%%%%%%%
\bibliographystyle{naturemag}
%\bibliography{bibliography}

\begin{thebibliography}{1}
\expandafter\ifx\csname url\endcsname\relax
  \def\url#1{\texttt{#1}}\fi
\expandafter\ifx\csname urlprefix\endcsname\relax\def\urlprefix{URL }\fi
\providecommand{\bibinfo}[2]{#2}
\providecommand{\eprint}[2][]{\url{#2}}

\bibitem{Note1}
\bibinfo{note}{TDVP equations for the thermofield purification of the density
  matrix may escape this fate\cite {hallam2019lyapunov} at least as far as
  local observations are concerned}.

\end{thebibliography}


\begin{thebibliography}{10}
\expandafter\ifx\csname url\endcsname\relax
  \def\url#1{\texttt{#1}}\fi
\expandafter\ifx\csname urlprefix\endcsname\relax\def\urlprefix{URL }\fi
\providecommand{\bibinfo}[2]{#2}
\providecommand{\eprint}[2][]{\url{#2}}

\bibitem{nahum2017quantum}
\bibinfo{author}{Nahum, A.}, \bibinfo{author}{Ruhman, J.},
  \bibinfo{author}{Vijay, S.} \& \bibinfo{author}{Haah, J.}
\newblock \bibinfo{title}{Quantum entanglement growth under random unitary
  dynamics}.
\newblock \emph{\bibinfo{journal}{Physical Review X}}
  \textbf{\bibinfo{volume}{7}}, \bibinfo{pages}{031016} (\bibinfo{year}{2017}).

\bibitem{li2018quantum}
\bibinfo{author}{Li, Y.}, \bibinfo{author}{Chen, X.} \&
  \bibinfo{author}{Fisher, M.~P.}
\newblock \bibinfo{title}{Quantum zeno effect and the many-body entanglement
  transition}.
\newblock \emph{\bibinfo{journal}{Physical Review B}}
  \textbf{\bibinfo{volume}{98}}, \bibinfo{pages}{205136}
  (\bibinfo{year}{2018}).

\bibitem{nahum2018operator}
\bibinfo{author}{Nahum, A.}, \bibinfo{author}{Vijay, S.} \&
  \bibinfo{author}{Haah, J.}
\newblock \bibinfo{title}{Operator spreading in random unitary circuits}.
\newblock \emph{\bibinfo{journal}{Physical Review X}}
  \textbf{\bibinfo{volume}{8}}, \bibinfo{pages}{021014} (\bibinfo{year}{2018}).

\bibitem{von2018operator}
\bibinfo{author}{von Keyserlingk, C.}, \bibinfo{author}{Rakovszky, T.},
  \bibinfo{author}{Pollmann, F.} \& \bibinfo{author}{Sondhi, S.}
\newblock \bibinfo{title}{Operator hydrodynamics, otocs, and entanglement
  growth in systems without conservation laws}.
\newblock \emph{\bibinfo{journal}{Physical Review X}}
  \textbf{\bibinfo{volume}{8}}, \bibinfo{pages}{021013} (\bibinfo{year}{2018}).

\bibitem{chan2019unitary}
\bibinfo{author}{Chan, A.}, \bibinfo{author}{Nandkishore, R.~M.},
  \bibinfo{author}{Pretko, M.} \& \bibinfo{author}{Smith, G.}
\newblock \bibinfo{title}{Unitary-projective entanglement dynamics}.
\newblock \emph{\bibinfo{journal}{Physical Review B}}
  \textbf{\bibinfo{volume}{99}}, \bibinfo{pages}{224307}
  (\bibinfo{year}{2019}).

\bibitem{li2019measurement}
\bibinfo{author}{Li, Y.}, \bibinfo{author}{Chen, X.} \&
  \bibinfo{author}{Fisher, M.~P.}
\newblock \bibinfo{title}{Measurement-driven entanglement transition in hybrid
  quantum circuits}.
\newblock \emph{\bibinfo{journal}{Physical Review B}}
  \textbf{\bibinfo{volume}{100}}, \bibinfo{pages}{134306}
  (\bibinfo{year}{2019}).

\bibitem{szyniszewski2019entanglement}
\bibinfo{author}{Szyniszewski, M.}, \bibinfo{author}{Romito, A.} \&
  \bibinfo{author}{Schomerus, H.}
\newblock \bibinfo{title}{Entanglement transition from variable-strength weak
  measurements}.
\newblock \emph{\bibinfo{journal}{Physical Review B}}
  \textbf{\bibinfo{volume}{100}}, \bibinfo{pages}{064204}
  (\bibinfo{year}{2019}).

\bibitem{misra1977zeno}
\bibinfo{author}{Misra, B.} \& \bibinfo{author}{Sudarshan, E.~G.}
\newblock \bibinfo{title}{The zeno's paradox in quantum theory}.
\newblock \emph{\bibinfo{journal}{Journal of Mathematical Physics}}
  \textbf{\bibinfo{volume}{18}}, \bibinfo{pages}{756--763}
  (\bibinfo{year}{1977}).

\bibitem{szyniszewski2020universality}
\bibinfo{author}{Szyniszewski, M.}, \bibinfo{author}{Romito, A.} \&
  \bibinfo{author}{Schomerus, H.}
\newblock \bibinfo{title}{Universality of entanglement transitions from
  stroboscopic to continuous measurements}.
\newblock \emph{\bibinfo{journal}{Physical review letters}}
  \textbf{\bibinfo{volume}{125}}, \bibinfo{pages}{210602}
  (\bibinfo{year}{2020}).

\bibitem{li2020conformal}
\bibinfo{author}{Li, Y.}, \bibinfo{author}{Chen, X.}, \bibinfo{author}{Ludwig,
  A.~W.} \& \bibinfo{author}{Fisher, M.}
\newblock \bibinfo{title}{Conformal invariance and quantum non-locality in
  hybrid quantum circuits}.
\newblock \emph{\bibinfo{journal}{arXiv preprint arXiv:2003.12721}}
  (\bibinfo{year}{2020}).

\bibitem{chen2020emergent}
\bibinfo{author}{Chen, X.}, \bibinfo{author}{Li, Y.}, \bibinfo{author}{Fisher,
  M.~P.} \& \bibinfo{author}{Lucas, A.}
\newblock \bibinfo{title}{Emergent conformal symmetry in nonunitary random
  dynamics of free fermions}.
\newblock \emph{\bibinfo{journal}{Physical Review Research}}
  \textbf{\bibinfo{volume}{2}}, \bibinfo{pages}{033017} (\bibinfo{year}{2020}).

\bibitem{tang2020measurement}
\bibinfo{author}{Tang, Q.} \& \bibinfo{author}{Zhu, W.}
\newblock \bibinfo{title}{Measurement-induced phase transition: A case study in
  the nonintegrable model by density-matrix renormalization group
  calculations}.
\newblock \emph{\bibinfo{journal}{Physical Review Research}}
  \textbf{\bibinfo{volume}{2}}, \bibinfo{pages}{013022} (\bibinfo{year}{2020}).

\bibitem{zhang2020nonuniversal}
\bibinfo{author}{Zhang, L.} \emph{et~al.}
\newblock \bibinfo{title}{Nonuniversal entanglement level statistics in
  projection-driven quantum circuits}.
\newblock \emph{\bibinfo{journal}{Physical Review B}}
  \textbf{\bibinfo{volume}{101}}, \bibinfo{pages}{235104}
  (\bibinfo{year}{2020}).

\bibitem{zabalo2020critical}
\bibinfo{author}{Zabalo, A.} \emph{et~al.}
\newblock \bibinfo{title}{Critical properties of the measurement-induced
  transition in random quantum circuits}.
\newblock \emph{\bibinfo{journal}{Physical Review B}}
  \textbf{\bibinfo{volume}{101}}, \bibinfo{pages}{060301}
  (\bibinfo{year}{2020}).

\bibitem{gullans2020scalable}
\bibinfo{author}{Gullans, M.~J.} \& \bibinfo{author}{Huse, D.~A.}
\newblock \bibinfo{title}{Scalable probes of measurement-induced criticality}.
\newblock \emph{\bibinfo{journal}{Physical review letters}}
  \textbf{\bibinfo{volume}{125}}, \bibinfo{pages}{070606}
  (\bibinfo{year}{2020}).

\bibitem{gullans2020dynamical}
\bibinfo{author}{Gullans, M.~J.} \& \bibinfo{author}{Huse, D.~A.}
\newblock \bibinfo{title}{Dynamical purification phase transition induced by
  quantum measurements}.
\newblock \emph{\bibinfo{journal}{Physical Review X}}
  \textbf{\bibinfo{volume}{10}}, \bibinfo{pages}{041020}
  (\bibinfo{year}{2020}).

\bibitem{jian2020measurement}
\bibinfo{author}{Jian, C.-M.}, \bibinfo{author}{You, Y.-Z.},
  \bibinfo{author}{Vasseur, R.} \& \bibinfo{author}{Ludwig, A.~W.}
\newblock \bibinfo{title}{Measurement-induced criticality in random quantum
  circuits}.
\newblock \emph{\bibinfo{journal}{Physical Review B}}
  \textbf{\bibinfo{volume}{101}}, \bibinfo{pages}{104302}
  (\bibinfo{year}{2020}).

\bibitem{choi2020quantum}
\bibinfo{author}{Choi, S.}, \bibinfo{author}{Bao, Y.}, \bibinfo{author}{Qi,
  X.-L.} \& \bibinfo{author}{Altman, E.}
\newblock \bibinfo{title}{Quantum error correction in scrambling dynamics and
  measurement-induced phase transition}.
\newblock \emph{\bibinfo{journal}{Physical Review Letters}}
  \textbf{\bibinfo{volume}{125}}, \bibinfo{pages}{030505}
  (\bibinfo{year}{2020}).

\bibitem{bao2020theory}
\bibinfo{author}{Bao, Y.}, \bibinfo{author}{Choi, S.} \&
  \bibinfo{author}{Altman, E.}
\newblock \bibinfo{title}{Theory of the phase transition in random unitary
  circuits with measurements}.
\newblock \emph{\bibinfo{journal}{Physical Review B}}
  \textbf{\bibinfo{volume}{101}}, \bibinfo{pages}{104301}
  (\bibinfo{year}{2020}).

\bibitem{goto2020measurement}
\bibinfo{author}{Goto, S.} \& \bibinfo{author}{Danshita, I.}
\newblock \bibinfo{title}{Measurement-induced transitions of the entanglement
  scaling law in ultracold gases with controllable dissipation}.
\newblock \emph{\bibinfo{journal}{Physical Review A}}
  \textbf{\bibinfo{volume}{102}}, \bibinfo{pages}{033316}
  (\bibinfo{year}{2020}).

\bibitem{alba2021spreading}
\bibinfo{author}{Alba, V.} \& \bibinfo{author}{Carollo, F.}
\newblock \bibinfo{title}{Spreading of correlations in markovian open quantum
  systems}.
\newblock \emph{\bibinfo{journal}{Physical Review B}}
  \textbf{\bibinfo{volume}{103}}, \bibinfo{pages}{L020302}
  (\bibinfo{year}{2021}).

\bibitem{fuji2020measurement}
\bibinfo{author}{Fuji, Y.} \& \bibinfo{author}{Ashida, Y.}
\newblock \bibinfo{title}{Measurement-induced quantum criticality under
  continuous monitoring}.
\newblock \emph{\bibinfo{journal}{Physical Review B}}
  \textbf{\bibinfo{volume}{102}}, \bibinfo{pages}{054302}
  (\bibinfo{year}{2020}).

\bibitem{PhysRevLett.126.170602}
\bibinfo{author}{Alberton, O.}, \bibinfo{author}{Buchhold, M.} \&
  \bibinfo{author}{Diehl, S.}
\newblock \bibinfo{title}{Entanglement transition in a monitored free-fermion
  chain: From extended criticality to area law}.
\newblock \emph{\bibinfo{journal}{Phys. Rev. Lett.}}
  \textbf{\bibinfo{volume}{126}}, \bibinfo{pages}{170602}
  (\bibinfo{year}{2021}).
\newblock
  \urlprefix\url{https://link.aps.org/doi/10.1103/PhysRevLett.126.170602}.

\bibitem{HaegemanTDVP}
\bibinfo{author}{Haegeman, J.} \emph{et~al.}
\newblock \bibinfo{title}{Time-dependent variational principle for quantum
  lattices}.
\newblock \emph{\bibinfo{journal}{Physical Review Letters}}
  \textbf{\bibinfo{volume}{107}}, \bibinfo{pages}{070601}
  (\bibinfo{year}{2011}).

\bibitem{hallam2019lyapunov}
\bibinfo{author}{Hallam, A.}, \bibinfo{author}{Morley, J.} \&
  \bibinfo{author}{Green, A.~G.}
\newblock \bibinfo{title}{The lyapunov spectra of quantum thermalisation}.
\newblock \emph{\bibinfo{journal}{Nature communications}}
  \textbf{\bibinfo{volume}{10}}, \bibinfo{pages}{1--8} (\bibinfo{year}{2019}).

\bibitem{zwanzig1960ensemble}
\bibinfo{author}{Zwanzig, R.}
\newblock \bibinfo{title}{Ensemble method in the theory of irreversibility}.
\newblock \emph{\bibinfo{journal}{The Journal of Chemical Physics}}
  \textbf{\bibinfo{volume}{33}}, \bibinfo{pages}{1338--1341}
  (\bibinfo{year}{1960}).

\bibitem{Orus:2014zl}
\bibinfo{author}{Or{\'u}s, R.}
\newblock \bibinfo{title}{A practical introduction to tensor networks: Matrix
  product states and projected entangled pair states}.
\newblock \emph{\bibinfo{journal}{Annals of physics}}
  \textbf{\bibinfo{volume}{349}}, \bibinfo{pages}{117--158}
  (\bibinfo{year}{2014}).

\bibitem{schollwock2011density}
\bibinfo{author}{Schollw{\"o}ck, U.}
\newblock \bibinfo{title}{The density-matrix renormalization group in the age
  of matrix product states}.
\newblock \emph{\bibinfo{journal}{Annals of Physics}}
  \textbf{\bibinfo{volume}{326}}, \bibinfo{pages}{96--192}
  (\bibinfo{year}{2011}).

\bibitem{perez2006matrix}
\bibinfo{author}{Perez-Garcia, D.}, \bibinfo{author}{Verstraete, F.},
  \bibinfo{author}{Wolf, M.~M.} \& \bibinfo{author}{Cirac, J.~I.}
\newblock \bibinfo{title}{Matrix product state representations}.
\newblock \emph{\bibinfo{journal}{arXiv preprint quant-ph/0608197}}
  (\bibinfo{year}{2006}).

\bibitem{kamenev2011field}
\bibinfo{author}{Kamenev, A.}
\newblock \emph{\bibinfo{title}{Field theory of non-equilibrium systems}}
  (\bibinfo{publisher}{Cambridge University Press}, \bibinfo{year}{2011}).

\bibitem{sieberer2016keldysh}
\bibinfo{author}{Sieberer, L.~M.}, \bibinfo{author}{Buchhold, M.} \&
  \bibinfo{author}{Diehl, S.}
\newblock \bibinfo{title}{Keldysh field theory for driven open quantum
  systems}.
\newblock \emph{\bibinfo{journal}{Reports on Progress in Physics}}
  \textbf{\bibinfo{volume}{79}}, \bibinfo{pages}{096001}
  (\bibinfo{year}{2016}).

\bibitem{green2016feynman}
\bibinfo{author}{Green, A.}, \bibinfo{author}{Hooley, C.},
  \bibinfo{author}{Keeling, J.} \& \bibinfo{author}{Simon, S.}
\newblock \bibinfo{title}{Feynman path integrals over entangled states}.
\newblock \emph{\bibinfo{journal}{arXiv preprint arXiv:1607.01778}}
  (\bibinfo{year}{2016}).

\bibitem{verstraete2004matrix}
\bibinfo{author}{Verstraete, F.}, \bibinfo{author}{Garcia-Ripoll, J.~J.} \&
  \bibinfo{author}{Cirac, J.~I.}
\newblock \bibinfo{title}{Matrix product density operators: Simulation of
  finite-temperature and dissipative systems}.
\newblock \emph{\bibinfo{journal}{Physical review letters}}
  \textbf{\bibinfo{volume}{93}}, \bibinfo{pages}{207204}
  (\bibinfo{year}{2004}).

\bibitem{cui2015variational}
\bibinfo{author}{Cui, J.}, \bibinfo{author}{Cirac, J.~I.} \&
  \bibinfo{author}{Banuls, M.~C.}
\newblock \bibinfo{title}{Variational matrix product operators for the steady
  state of dissipative quantum systems}.
\newblock \emph{\bibinfo{journal}{Physical review letters}}
  \textbf{\bibinfo{volume}{114}}, \bibinfo{pages}{220601}
  (\bibinfo{year}{2015}).

\bibitem{weimer2019simulation}
\bibinfo{author}{Weimer, H.}, \bibinfo{author}{Kshetrimayum, A.} \&
  \bibinfo{author}{Or{\'u}s, R.}
\newblock \bibinfo{title}{Simulation methods for open quantum many-body
  systems}.
\newblock \emph{\bibinfo{journal}{arXiv preprint arXiv:1907.07079}}
  (\bibinfo{year}{2019}).

\bibitem{daley2009atomic}
\bibinfo{author}{Daley, A.}, \bibinfo{author}{Taylor, J.},
  \bibinfo{author}{Diehl, S.}, \bibinfo{author}{Baranov, M.} \&
  \bibinfo{author}{Zoller, P.}
\newblock \bibinfo{title}{Atomic three-body loss as a dynamical three-body
  interaction}.
\newblock \emph{\bibinfo{journal}{Physical review letters}}
  \textbf{\bibinfo{volume}{102}}, \bibinfo{pages}{040402}
  (\bibinfo{year}{2009}).

\bibitem{daley2014quantum}
\bibinfo{author}{Daley, A.~J.}
\newblock \bibinfo{title}{Quantum trajectories and open many-body quantum
  systems}.
\newblock \emph{\bibinfo{journal}{Advances in Physics}}
  \textbf{\bibinfo{volume}{63}}, \bibinfo{pages}{77--149}
  (\bibinfo{year}{2014}).

\bibitem{bonnes2014superoperators}
\bibinfo{author}{Bonnes, L.} \& \bibinfo{author}{L{\"a}uchli, A.~M.}
\newblock \bibinfo{title}{Superoperators vs. trajectories for matrix product
  state simulations of open quantum system: a case study}.
\newblock \emph{\bibinfo{journal}{arXiv preprint arXiv:1411.4831}}
  (\bibinfo{year}{2014}).

\bibitem{strathearn2018efficient}
\bibinfo{author}{Strathearn, A.}, \bibinfo{author}{Kirton, P.},
  \bibinfo{author}{Kilda, D.}, \bibinfo{author}{Keeling, J.} \&
  \bibinfo{author}{Lovett, B.~W.}
\newblock \bibinfo{title}{Efficient non-markovian quantum dynamics using
  time-evolving matrix product operators}.
\newblock \emph{\bibinfo{journal}{Nature communications}}
  \textbf{\bibinfo{volume}{9}}, \bibinfo{pages}{1--9} (\bibinfo{year}{2018}).

\bibitem{FergusThesis}
\bibinfo{author}{Barratt, F.}
\newblock \emph{\bibinfo{title}{The dynamics and control of quantum information
  out of equilibrium}}.
\newblock Ph.D. thesis, \bibinfo{school}{Kings College London}
  (\bibinfo{year}{2021}).

\bibitem{Note1}
\bibinfo{note}{TDVP equations for the thermofield purification of the density
  matrix may escape this fate\cite {hallam2019lyapunov} at least as far as
  local observations are concerned}.

\bibitem{banuls2011strong}
\bibinfo{author}{Ba{\~n}uls, M.~C.}, \bibinfo{author}{Cirac, J.~I.} \&
  \bibinfo{author}{Hastings, M.~B.}
\newblock \bibinfo{title}{Strong and weak thermalization of infinite
  nonintegrable quantum systems}.
\newblock \emph{\bibinfo{journal}{Physical review letters}}
  \textbf{\bibinfo{volume}{106}}, \bibinfo{pages}{050405}
  (\bibinfo{year}{2011}).

\bibitem{leviatan2017quantum}
\bibinfo{author}{Leviatan, E.}, \bibinfo{author}{Pollmann, F.},
  \bibinfo{author}{Bardarson, J.~H.}, \bibinfo{author}{Huse, D.~A.} \&
  \bibinfo{author}{Altman, E.}
\newblock \bibinfo{title}{Quantum thermalization dynamics with matrix-product
  states}.
\newblock \emph{\bibinfo{journal}{arXiv preprint arXiv:1702.08894}}
  (\bibinfo{year}{2017}).

\bibitem{leggett1987dynamics}
\bibinfo{author}{Leggett, A.~J.} \emph{et~al.}
\newblock \bibinfo{title}{Dynamics of the dissipative two-state system}.
\newblock \emph{\bibinfo{journal}{Reviews of Modern Physics}}
  \textbf{\bibinfo{volume}{59}}, \bibinfo{pages}{1} (\bibinfo{year}{1987}).

\bibitem{Florens2010quantum}
\bibinfo{author}{Florens, S.}, \bibinfo{author}{Venturelli, D.} \&
  \bibinfo{author}{Narayanan, R.}
\newblock \bibinfo{title}{Quantum phase transition in the spin boson model}.
\newblock In \emph{\bibinfo{booktitle}{Quantum Quenching, Annealing and
  Computation}}, \bibinfo{pages}{145--162} (\bibinfo{publisher}{Springer},
  \bibinfo{year}{2010}).

\bibitem{barratt2020dissipative}
\bibinfo{author}{Barratt, F.} \emph{et~al.}
\newblock \bibinfo{title}{Dissipative failure of adiabatic quantum transport as
  a dynamical phase transition}.
\newblock \emph{\bibinfo{journal}{arXiv preprint arXiv:2012.15212}}
  (\bibinfo{year}{2020}).

\bibitem{crowley2014quantum}
\bibinfo{author}{Crowley, P.}, \bibinfo{author}{{\DJ}uri{\'c}, T.},
  \bibinfo{author}{Vinci, W.}, \bibinfo{author}{Warburton, P.} \&
  \bibinfo{author}{Green, A.}
\newblock \bibinfo{title}{Quantum and classical dynamics in adiabatic
  computation}.
\newblock \emph{\bibinfo{journal}{Physical Review A}}
  \textbf{\bibinfo{volume}{90}}, \bibinfo{pages}{042317}
  (\bibinfo{year}{2014}).

\bibitem{zwanzig1961memory}
\bibinfo{author}{Zwanzig, R.}
\newblock \bibinfo{title}{Memory effects in irreversible thermodynamics}.
\newblock \emph{\bibinfo{journal}{Physical Review}}
  \textbf{\bibinfo{volume}{124}}, \bibinfo{pages}{983} (\bibinfo{year}{1961}).

\bibitem{berta2017thermal}
\bibinfo{author}{Berta, M.}, \bibinfo{author}{Brandao, F.~G.},
  \bibinfo{author}{Haegeman, J.}, \bibinfo{author}{Scholz, V.~B.} \&
  \bibinfo{author}{Verstraete, F.}
\newblock \bibinfo{title}{Thermal states as convex combinations of matrix
  product states}.
\newblock \emph{\bibinfo{journal}{arXiv preprint arXiv:1709.07423}}
  (\bibinfo{year}{2017}).
  
\bibitem{frenkel1934wave}
\bibinfo{author}{Frenkel, J.}
\newblock \emph{\bibinfo{title}{Wave mechanics, advanced general theory}},
  vol.~\bibinfo{volume}{1} (\bibinfo{publisher}{Oxford}, \bibinfo{year}{1934}).

\bibitem{kamenev2002keldysh}
\bibinfo{author}{Kamenev, A.}
\newblock \bibinfo{title}{Keldysh and doi-peliti techniques for
  out-of-equilibrium systems}.
\newblock In \emph{\bibinfo{booktitle}{Strongly Correlated Fermions and Bosons
  in Low-Dimensional Disordered Systems}}, \bibinfo{pages}{313--340}
  (\bibinfo{publisher}{Springer}, \bibinfo{year}{2002}).

\bibitem{crowley2016anisotropic}
\bibinfo{author}{Crowley, P.~J.} \& \bibinfo{author}{Green, A.}
\newblock \bibinfo{title}{Anisotropic landau-lifshitz-gilbert models of
  dissipation in qubits}.
\newblock \emph{\bibinfo{journal}{Physical Review A}}
  \textbf{\bibinfo{volume}{94}}, \bibinfo{pages}{062106}
  (\bibinfo{year}{2016}).

\bibitem{haegeman2016unifying}
\bibinfo{author}{Haegeman, J.}, \bibinfo{author}{Lubich, C.},
  \bibinfo{author}{Oseledets, I.}, \bibinfo{author}{Vandereycken, B.} \&
  \bibinfo{author}{Verstraete, F.}
\newblock \bibinfo{title}{Unifying time evolution and optimization with matrix
  product states}.
\newblock \emph{\bibinfo{journal}{Physical Review B}}
  \textbf{\bibinfo{volume}{94}}, \bibinfo{pages}{165116}
  (\bibinfo{year}{2016}).

\end{thebibliography}

%%%%%%%%%%%%%%%%%%%%%%%%%%%%%%%%%%%%%%%%

\appendix

\section{Haar averaging on the MPS manifold}
\label{app:HaarAverage}
%%%%%%%%%%%%%%%%%%
%\input{AppendixC.tex}
In the left canonical form, MPS of left bond order $D$ and local Hilbert space dimension $D$ are given by $SU(dD)$ isometries\cite{Orus:2014zl}. In the case of a finite chain of length $L$, the left bond order at the $n^{th}$ site $D_n=d^n$ up to the maximum bond order at site $n= \log_d D_{\hbox{max}}$. The bond order remains $D_{\hbox{max}}$ up to site $L-1+\log_d D_{\hbox{max}}$ beyond which it reduces as $D_n=d^{(n-L+1)}$.

The thermal expectation of an operator can be calculated by a Boltzmann-weighted Haar average over this variational manifold. The average of an operator $\hat O$ is given by 
\begin{eqnarray}
\langle \langle \hat  O \rangle \rangle
&=&
\frac{
\int \prod_n DA_n \langle \phi | \hat O | \phi \rangle 
\exp \left[ - \beta \langle \phi| \hat H | \phi \rangle   \right]
}{
\int \prod_n DA_n 
\exp \left[ - \beta \langle \phi| \hat H | \phi \rangle   \right]
}.
\end{eqnarray}
The expectation of the Hamiltonian $ \langle \phi| \hat H | \phi \rangle$ and the operator $ \langle \phi | \hat O | \phi \rangle $ are calculated by usual MPS techniques. 
In practice, we carry out the integrals by by sampling a Haar random distribution of isometric MPS tensors; $A^\sigma_{ij} \equiv U_{i \otimes \sigma, j\otimes 1} \in SU(dD_n)$ . These are obtained by a QR decomposition of a tensor with elements drawn randomly from a normal distribution. 

\begin{figure}[h]
\includegraphics[width=0.47\textwidth]{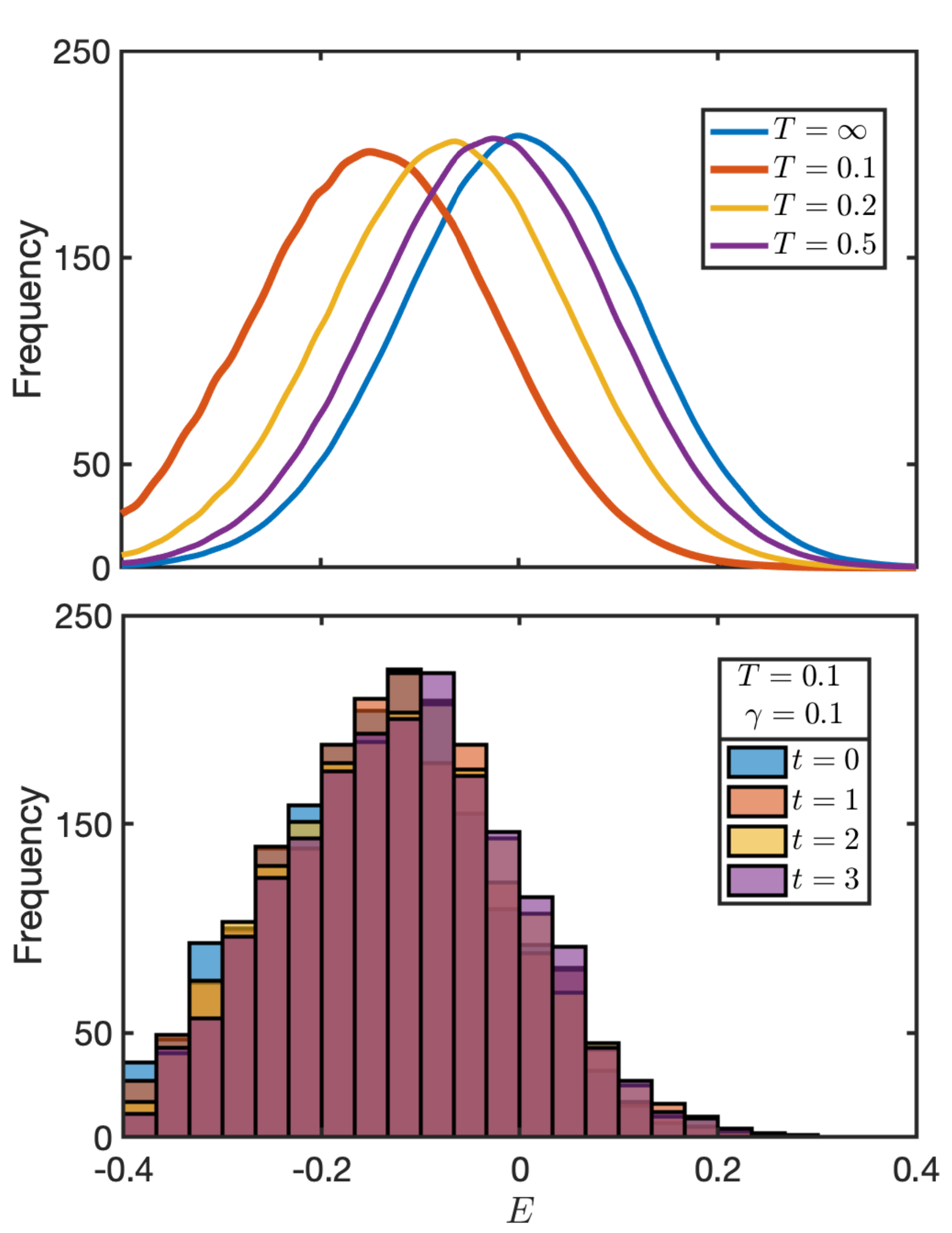}
\caption{
{\it Density of occupied states at finite-temperature for a length 15 chain at bond order} $D_{\hbox{max}}=2$: a) The density of occupied states is computed for the Hamiltonian 
Eq.(2)
%(\ref{eq:Hamiltonian}) 
with $J=1 $ $g=-1.05$, and $h=0.5$ from a sample of Haar random distributed isometric MPS initial states. At $T=\infty$ --- an unweighted Haar average --- the majority of the states are in the middle of the spectrum and are highly entangled. A Boltzmann weight shifts this distribution to lower energy. Low-entanglement states are a subset of measure zero in the thermodynamic limit. b) The Boltzmann-weighted Haar average is a fixed point of the TDVP-Langevin equation. Here we compare distributions obtained after evolving with the Hamiltonian for $t=1$, $2$, and $3$, with $\gamma = 0.1$ and $T=0.1$. The distribution for $T=0.1$ is shown in bold in a), we see the Langevin evolution preserves this distribution.
} 
\label{appfig:BoltzmannHaar}
\end{figure}

Fig.~\ref{appfig:BoltzmannHaar} shows this Boltzmann-weighted Haar distribution as a function of energy at different temperatures. These thermal distributions are fixed-points of the dynamics described by Eq.(1)%(\ref{eq:TDVPLangevin}).

\section{Deriving the TDVP Langevin Equation}
\label{app:TDVPLangevin}
%%%%%%%%%%%%%%%%%%
%\input{AppendixA.tex}
Here we outline how the MPS TDVP Langevin equation can be obtained. We follow two separate routes: a heuristic route using a solution of the Schr\'odinger equation for the bath and system, and constructing a Langevin limit of the Keldysh path integral constructed over MPS. 
 
{\it Bath Model:} We model the bath as a collection of independent non-interacting harmonic oscillators. These are coupled to local system operators $\hat F_n$ at site $n$ by their displacements. The Hamiltonian for the bath and its coupling to the system are given by
\begin{eqnarray}
\hat H_{bath}
&=&
\sum_n \sum_\alpha 
 \hbar \omega_\alpha \left( \hat a^\dagger_{n,\alpha}  \hat a_{n,\alpha} +1/2 \right),
%\sum_n \sum_\alpha 
%\left( 
%\frac{\hat p_{n,\alpha}^2}{2 m_{\alpha}} 
%+ \frac{1}{2} m_{\alpha} \omega_{\alpha}^2 \hat x^2_{n,\alpha}
%\right),
\nonumber\\
\hat H_{I}
&=& 
- \sum_n \sum_\alpha \lambda_\alpha \left( \hat a^\dagger_{n,\alpha}+\hat a_{n,\alpha}  \right) \hat F_n,
%- \sum_n \sum_\alpha \lambda_\alpha \hat x_{n,\alpha} \hat F_n,
\end{eqnarray}
respectively. The index $\alpha$ labels the different oscillator modes at the site $n$. The distribution and temperature of the oscillator modes is assumed to be the same at each site. Moreover, we assume no back reaction of the system on the bath, so that the bath distribution remains equilibrium. This is a subtle assumption - the bath must be non-linear in order to thermalise energy absorbed from the system, but these non-linearities must operate on timescales such that the bath's effect on the system is the same as independent oscillators. The assumptions are standard - but non-trivial - and permit the simple manipulations that follow.

{\it A heuristic derivation of the TDVP Langevin equation} can be made in the spirit of the Frenkel principle for deriving the TDVP equations \cite{frenkel1934wave}. The state of the system and bath is parametrized as $| \psi(\bf{z}) \rangle \otimes_{n, \alpha} |\phi_{n, \alpha} \rangle$ where ${\bf z}$ corresponds to some set of variational parameters of the system and $\phi_{n,\alpha}$ are coherent state parameters of the $\alpha$-oscillator on site $n$. The time derivative of the wavefunction in the Schr\"odinger equation for the system and bath is expanded in a chain rule over the system and bath parameters:
\begin{eqnarray*}
& &
\dot {\bf z} |\partial_{z_i} \psi \rangle \otimes_{n,\alpha} |\phi_{n,\alpha} \rangle 
+| \psi \rangle \sum_{m,\beta} \dot \phi_{m,\beta} \partial_{\phi_{m,\beta}} \left( \otimes_{n,\alpha} |\phi_{n,\alpha} \rangle \right)
\\
&\approx&
- i \hat H |\psi \rangle \otimes_{n,\alpha} | \phi_\alpha \rangle, 
\end{eqnarray*}
where the inequality is because the dynamics might take the state outside of the variational manifold. Taking an inner product with 
$\langle \partial_{z_i} \psi |\otimes_{n,\alpha} \langle \phi_{n,\alpha} |$ 
allows us to obtain an equation of motion for the system in the presence of the bath, and an inner product with 
$\langle \psi | \; \partial_{\phi_{o,\gamma}} \left(  \otimes_{n,\alpha} \langle \phi_{n,\alpha} | \right) $ 
allows us to obtain an equation of motion for the bath in the presence of the system:
\begin{eqnarray}
& &i\langle \partial_{z_i} \psi | \partial_{z_j} \psi \rangle \dot z_j
 =\langle \partial_{z_i} \psi | \hat H_S | \psi \rangle
\nonumber\\
& &
+\sum_n \langle \partial_{z_i} \psi | \hat F_{n} | \psi \rangle
\sum_\alpha \tilde \lambda_\alpha ( \bar \phi_{n,\alpha}(t)+ \phi_{n,\alpha}(t))
\nonumber\\
& &i \dot \phi_{n,\alpha} 
=
\omega_\alpha + \lambda_\alpha \langle \psi | \hat F_{n} | \psi \rangle
\end{eqnarray}
%
%In order to obtain these results, we use $\hat a_{n \alpha} | \phi_{\n,\alpha} \rangle = \phi_{n, \alpha} | \phi_{\n,\alpha} \rangle $ together with the normalisation $\langle \phi | \phi \rangle = e^{\bar \phi \phi}$.%
The equalities are attained since the inner products with the tangent vectors project the Hamiltonian evolution back onto the variational manifold. These equations correspond to the usual TDVP equations with addition of coupling between the system and bath.

The remaining steps involve integrating the equation of motion for the bath degrees of freedom and substituting back into the equation of motion for the system. The formal solution of the equation of motion of the bath degrees of freedom is
$$
\phi_{n,\alpha}(t) =\phi_{n,\alpha}(0) e^{-i \omega_\alpha t} - \int_0^\infty D^R(t'-t) \langle \psi | \hat F_n | \psi \rangle dt'
$$
where we have identified the retarded bath correlator $D^R(t'-t)= - i \Theta(t-t') \langle \hat a^\dagger(t') \hat a(t) \rangle = i e^{i \omega(t'-t)}$. The assumptions of no back reaction of the system on the bath are taken account of by treating the terms $\eta_{n}(t) =\sum_\alpha \phi_{n,\alpha}(0) e^{-i \omega_\alpha t} $ as a stochastic random field with variance appropriate to the thermal distribution. After further identifying $\partial_t \Gamma(t) = D^R(t)$ the system of equation of motion can be put in the form
\begin{widetext}
\begin{eqnarray}
& &i\langle \partial_{A_i} \psi | \partial_{A_j} \psi \rangle \dot A_j
 =\langle \partial_{A_i} \psi | \hat H_S | \psi \rangle
%\nonumber\\
%& &
+\sum_n \langle \partial_{A_i} \psi | \hat F_{n} | \psi \rangle
\left(
\int dt' \Gamma(t-t')  \langle \partial_{A_j} \psi | \hat F_{n} | \psi \rangle \dot A_j
+\eta(t)
\right)
\label{eq:NonMarkovianTDVPLangevin}
\end{eqnarray}
\end{widetext}
the Markovian limit of which recovers Eq.(1).
%(\ref{eq:TDVPLangevin})

{\it A formal derivation from the Keldysh path integral} can also be made. We follow the approach described in Ref.\cite{kamenev2002keldysh,kamenev2011field,crowley2016anisotropic}  constructing a Keldysh path integral for the density matrix and integrate out the bath in an appropriate limit to obtain a Langevin equation. Our main modification is to construct the path integral over MPS following Ref.\cite{green2016feynman}.

The initial density matrix is assumed to factorise into density matrices for the system and bath as $\hat \rho \otimes \hat \rho_{bath}$ - the bath being in thermal equilibrium and its distribution assumed to be unchanged in time. This evolves to 
$ \bar T e^{- i \int_0^t dt' (\hat H+ \hat H_{int} + \hat H_{bath})} \hat \rho \otimes \hat \rho_{bath}   T e^{- i \int_0^t dt' (\hat H+ \hat H_{int} + \hat H_{bath})} $ at time $t$. Construction of the Keldysh path integral proceeds by dividing up the time-ordered ($T$) and anti-time-ordered ($\bar T$) exponentials into many infinitessimal evolutions and inserting resolutions of the identity over coherent state variables for the bath and using a Haar measure over MPS states for the system\cite{green2016feynman}:
\begin{eqnarray*}
{\bm 1} 
&=&
\int \prod_n d\bar \phi_n(t) d \phi_n(t) e^{- \sum_n \bar \phi_n \phi_n} \otimes_n | \bar \phi_n \rangle \langle \phi_n |
\\
{\bm 1} 
&=& 
\int  DA(t) | A \rangle \langle A | 
\end{eqnarray*}
Here $DA$ symbolises an integral over all tensors in the MPS chain using the Haar measure introduced in Ref.\cite{green2016feynman}. A Keldysh rotation transforms from the fields  on the time-ordered contour ($+$) and 
%$A^-$, $\bar \phi^-, \phi^-$ 
 the anti time-ordered contour ($-$) to the sum and difference between them $A^\pm= A^{cl} \pm A^q$, $\phi^\pm= \phi^{cl}\pm \phi^q$, known as the classical and quantum components of the quantum field. 
 \begin{widetext}
 These manipulations give the following path integral for the evolution kernal of the density matrix:
\begin{eqnarray*}
K(t)
&=&
\int DA D (\bar \phi, \phi)e^{S[A^c,A^q,\phi^c,\phi^q]}
\\
S 
&=&
 S[A^{cl}+A^q]-S[A^{cl}-A^q]
%\\
%& &
+ \int dt \Bigg[ \sum_{n,\alpha} \lambda_\alpha 
\left( F^q_n,F^{cl}_n \right)
\left( \begin{array}{c} \phi^{cl}_{n,\alpha}+\bar \phi^{cl}_{n,\alpha} \\  \phi^{q}_{n,\alpha}+\bar \phi^{q}_{n,\alpha} \end{array} \right)
\Bigg]
\\
 & &
+ \int dt  \; dt' \Bigg[ \sum_{n,\alpha} 
\left(  \bar \phi^{cl}_{n,\alpha}(t) , \bar \phi^{q}_{n,\alpha}(t)  \right)
\begin{pmatrix}
0                 & [D_\alpha^A]^{-1}(t-t') \\
 [D_\alpha^R]^{-1}(t-t') & [D_\alpha^{-1}]^K(t-t') \\
  \end{pmatrix}(t-t')
 \left( \begin{array}{c} \phi^{cl}_{n,\alpha}(t') \\  \phi^{q}_{n,\alpha}(t') \end{array} \right)
 \Bigg]
\end{eqnarray*}
where $D_\alpha^A$, $D_\alpha^R$ and $D_\alpha^K$ are the advanced, retarded and Keldysh components of the bath Greens function. Consistent with our assumption of a thermal equilibrium bath and no back reaction, they are related by the fluctuation dissipation relation: $D^K_\alpha(\omega) = \coth (\omega /2T) [D^R_\alpha(\omega) - D^A_\alpha(\omega)]$ with $D^{R(A)}=1/(\omega \pm i \delta)$. $S[A]$ is the action of the system in the absence of coupling to the bath. 

The simple quadratic form of the bath action follows from our assumptions and modeling of its affects as independent harmonic oscillators. It enables one to integrate out the bath and, depending upon timescales\cite{Fergus}, to construct either a Lindblad or Langevin limit. We construct the latter limit in three steps. First the bath degrees of freedom are integrated out. The resulting dissipative contribution to the action has cross terms between classical and quantum components of the expectations of $\hat F$, and a term quadratic in the quantum component.,
$$
S_{diss}
=
\int dt \; dt'
\sum_n \left( F^{cl}_n , F^{q}_n  \right)(t)
\begin{pmatrix}
0                 & D^A(t-t') \\
 D^R(t-t') & D^K(t-t') \\
  \end{pmatrix}
 \left( \begin{array}{c} F^{cl}_{n} \\  F^{q}_{n,} \end{array} \right)(t'),
$$
where the bath propagators without indices indicate a sum over all modes, for example $D^R=\sum_\alpha \lambda^2_\alpha D^R_\alpha$. The quadratic term in $F^q$ is decoupled with a Hubbard-Stratonovich field $\eta(t) $ that ultimately will play the role of the stochastic noise field in Eq.(1).
%\ref{eq:TDVPLangevin}). 
The final trick to bring this integral to the Langevin form is to Taylor expand the action to linear order in the  in the quantum fields, $A^q$.
The result is a path integral over the MPS tensors $A^q$ and $A^{cl}$ and the noise field $\eta$:
\begin{eqnarray*}
K(t)
&=&
\int DA^q DA^{cl} D\eta e^{iS[A^c,A^q,\eta]}
\\
S 
&=&
 \int dt 
\sum_n A^q_n(t) \underbrace{\Bigg[
2 \delta S[A^{cl}]/\delta A_n^{cl}(t)+
  2 \sum_m \partial F^{cl}_m/\partial A_n^{cl}(t) 
  \left(  \int dt' \; D^R(t-t') F^{cl}_m(t') + \eta_m(t) \right) \Bigg]}_{\hbox{Eq.(1)}}
 \\
 & &
 - \int dt\; dt'  \sum_n \eta_n(t) [D^K]^{-1}(t-t') \eta_n(t') \Bigg].
\end{eqnarray*}
This is equivalent to the TDVP Langevin of 
%%%%%%%%%%%%%%%%%%%%%%%
Eq.(\ref{eq:NonMarkovianTDVPLangevin}); 
%%%%%%%%%%%%%%%%%%%%%%%
the quantum field $A^q$ plays the role of a Lagrange multiplier that imposes 
%%%%%%%%%%%%%%%%%%%%%%%
Eq.(\ref{eq:NonMarkovianTDVPLangevin}) 
%%%%%%%%%%%%%%%%%%%%%%%%
and the remaining term gives the bath correlations. The tensor indices have been suppressed for clarity in this expression.
\end{widetext}
To make the comparison with the TDVP Langevin equation, note that $ \delta S[A^{cl}]/\delta A_n^{cl}(t)=0$ recovers the usual TDVP equations for matrix product states. The additional terms correspond to the dissipative effects of the bath. These terms are non-local in the chain indices $n$ and $m$, despite our model of local independent baths. This is due to the potential long-ranged entanglement of the matrix product state. The long-range effects of the noise term reflect those already found in the usual TDVP equations, since the noise term arises from a random local potential. The non-locality of the friction term is more problematic and some insight is required to implement it efficiently.

\section{Implementing the TDVP Langevin Equation}
\label{app:TDVP}
%%%%%%%%%%%%%%%%%%
%\input{AppendixB.tex}
Here we outline how the TDVP Langevin equation, Eq.(1),
%(\ref{eq:TDVPLangevin}), 
can be implemented numerically for MPS. This equation comprises three parts. The first is the closed-system TDVP equation. This is implemented by standard means. The second is the random noise induced by the environment. This is essentially a time-dependent Hamiltonian term. We integrate it in a Stratonovich scheme. The final part is friction. Even in the Markovian limit, this term is generally spatially non-local --- however, a significant simplification can be acheived by working with purely local operators. 
%We employ a ... scheme in order to deal with it efficiently.

\noindent
{\it The TDVP equations for  MPS} can be written n the form 
\begin{eqnarray}
\langle \partial_i \psi | \partial_j \psi \rangle \dot X_j
&=&
-i \langle \partial_i \psi | \hat H | \psi \rangle
\nonumber\\
\Rightarrow \;\;\;\; 
\langle \partial_{A_n} \psi | \partial_{A_m} \psi \rangle \dot A_m
&=&
-i \langle \partial_{A_n} \psi | \hat H | \psi \rangle 
\label{eq:TDVP}
\end{eqnarray}
where we have suppressed the tensor indices of $A$ for clarity, retaining only the site index. The solution of this equation is well established. A judicious choice of gauge fixing for the tangent vectors to the MPS manifold puts the Gramm matrix $\langle \partial_{A_n} \psi | \partial_{A_m} \psi \rangle$ in a diagonal form.  Various algorithms for evaluating the TDVP equations for finite systems exist, in this report we have used a modification of the method introduced in \cite{haegeman2016unifying}. A single time-step of the algorithm is achieved by sweeping through the system from right to left and applying a unitary rotation to the local variables on each site $A_n(t+\delta t)=e^{iH_{eff}}A_n(t)$, followed by repeating this process by sweeping from left to right. 

 \noindent
{\it The Noise } contribution to the TDVP Langevin equation for MPS can be written in the form
\begin{eqnarray}
& &
-i \sum_m \langle \partial_i \psi | \hat F_m | \psi \rangle \eta(t)
\nonumber\\
& \Rightarrow&
-i \sum_m \langle \partial_{A_n} \psi | \hat F_m | \psi \rangle \eta_m(t)
\label{eq:Noise}
\end{eqnarray}
This evidently takes the same form as the right hand side of 
%%%%%%%%%
Eq.(\ref{eq:TDVP})
%%%%%%%%%% 
and no substantial modification to the TDVP algorithm is required. At the beginning of each timestep, $\eta_m(t)$ is chosen by sampling from a normal distribution with mean zero and variance $2 \delta t \gamma T$.

 \noindent
{\it Friction:}
The friction term can be written in the Markovian limit and  in terms of MPS tensors as follows:
\begin{eqnarray}
& &
-i \sum_m \gamma \frac{\langle \psi | \hat F_m | \psi \rangle }{dt}  \langle \partial_i \psi | \hat F_m  | \psi \rangle,
%\nonumber\\
\label{eq:Friction}
\end{eqnarray}
where
\begin{equation}
\frac{\langle \psi | \hat F_m | \psi \rangle }{dt}=\langle \psi | \hat F_m |\partial_{A_n} \psi \rangle \dot{A}_n+  \langle\partial_{\bar{A}_n} \psi | \hat F_m |\psi \rangle \dot{\bar{A}}_n.
\label{eq:Friction_tderivative}
\end{equation}
\begin{widetext}
Eq.(\ref{eq:Friction_tderivative}) 
can be evaluated by substituting in 
Eq.(\ref{eq:TDVP}) 
introducing the poisson bracket notation, $\{O_1,O_2\}=i\langle \psi | \hat O_1 |\partial_{A_n} \psi \rangle \langle \partial_{\bar{A}_n}\psi | \hat O_2 |\psi \rangle-i \langle \psi | \hat O_2 |\partial_{A_n} \psi \rangle \langle \partial_{\bar{A}_n}\psi | \hat O_1 |\psi \rangle$,
\begin{equation}
\begin{split}
& (\delta_{m,n}+\sum_n \gamma \{F_m,F_n \})\frac{\langle \psi | \hat F_n | \psi \rangle }{dt}=\{F_m,H\}+\sum_n \{F_m,F_n\} \eta_n(t) \\
& \rightarrow \frac{\langle \psi | \hat F_n | \psi \rangle }{dt} = (\mathbb{I}+\gamma \mathbb{F})^{-1}\left(\{F_m,H\}+\sum_n \{F_m,F_n\} \eta_n(t) \right)
\end{split}
\label{eq:Friction_expression}
\end{equation}
where $\mathbb{F}_{ij}=\{F_i,F_j \}$. Evaluating 
Eq.(\ref{eq:Friction_expression}) 
exactly for arbitrary operators $\hat{F}_n$ is quite numerically inefficient, scaling quadratically in the number of noise operators. Moreover, it is inconsistent with the site by site sweep algorithm introduced above for the TDVP equations. Fortunately, 
Eq.(\ref{eq:Friction_expression})
is substantially simplified in the case of single-site, local noise fields. For local fields $\{F_i,F_j \}$ is only non-zero provided the two operators are located on the same site of the system and so $\mathbb{F}$ becomes a simple, block diagonal matrix. 
\end{widetext}

The modified TDVP algorithm therefore works as follows: Before each sweep through the system $\eta_m(t)$ is sampled from a normal distribution with mean $0$ and variance $2 \delta t \gamma T$. The Hamiltonian and noise terms are then used to calculate $\frac{\langle \psi | \hat F_n | \psi \rangle }{dt}$ for all noise operators $F_n$ using 
Eq.(\ref{eq:Friction_expression}).  
The noise and friction terms are then combine with the Hamiltonian using 
Eq.(\ref{eq:TDVP}) 
and the state is evolved using the standard TDVP algorithm.

\end{document}